\documentclass[aps, pra, reprint, showpacs, superscriptaddress, floatfix, longbibliography]{revtex4-1}
\pdfoutput=1
\usepackage{braket}
\usepackage{amsmath,dsfont, bm}
\usepackage{amsfonts} 
\usepackage{dsfont}
\usepackage{hyperref}
\hypersetup{pdfstartview=} 
\usepackage{graphicx}
\usepackage[dvipsnames]{xcolor}
\usepackage{color}
\usepackage[normalem]{ulem}

\DeclareMathOperator{\Tr}{Tr}
\DeclareMathOperator{\eig}{eig}
\DeclareMathOperator{\lik}{\mathcal{L}}
\newcommand{\Prob}{\text{Prob}}

\newif\ifcmnt
\cmnttrue 
\ifcmnt
    \providecommand{\aucmnt}[1]{#1}
    
\else
    \providecommand{\aucmnt}[1]{}
    
\fi

\begin{document}
\title{Joint Quantum-State and Measurement Tomography with Incomplete Measurements}
\date{\today}
\author{Adam C. Keith}
\email{adam.keith@colorado.edu}
\affiliation{Applied and Computational Mathematics Division, National Institute of Standards and Technology, Boulder, 
Colorado, 80305, USA}
\affiliation{Department of Physics, University of Colorado, Boulder, Colorado, 80309, USA}
\author{Charles H. Baldwin}
\affiliation{Applied and Computational Mathematics Division, National Institute of Standards and Technology, Boulder, 
Colorado, 80305, USA}
\author{Scott Glancy}
\email{sglancy@nist.gov}
\affiliation{Applied and Computational Mathematics Division, National Institute of Standards and Technology, Boulder, 
Colorado, 80305, USA}
\author{E. Knill}
\affiliation{Applied and Computational Mathematics Division, National Institute of Standards and Technology, Boulder, 
Colorado, 80305, USA}
\affiliation{Center for Theory of Quantum Matter, University of Colorado, Boulder, Colorado 80309, USA}

\begin{abstract}
  Estimation of quantum states and measurements is crucial for the
  implementation of quantum information protocols. The standard method
  for each is quantum tomography.  However, quantum tomography suffers
  from systematic errors caused by imperfect knowledge of the
  system. We present a procedure to simultaneously characterize
  quantum states and measurements that mitigates systematic errors by
  use of a single high-fidelity state preparation and a limited set of
  high-fidelity unitary operations. Such states and operations are
  typical of many state-of-the-art systems. For this situation we
  design a set of experiments and an optimization algorithm that
  alternates between maximizing the likelihood with respect to the
  states and measurements to produce estimates of each.  In some
  cases, the procedure does not enable unique estimation of the
  states.  For these cases, we show how one may identify a set of
  density matrices compatible with the measurements and use a
  semi-definite program to place bounds on the state's expectation
  values.  We demonstrate the procedure on data from a simulated
  experiment with two trapped ions.
\end{abstract}


\maketitle

\section{Introduction}

Recent experiments have demonstrated high-fidelity unitary operations
in various platforms for quantum information processing; for examples
see Refs.~\cite{Gaebler16, Nichol17, Sheldon16, Casparis16, Barends14,
  Ballance16}. Even in the most advanced systems, some operations are
harder to accomplish and have significantly lower fidelity than
others.  For example, in the systems reported in
Refs.~\cite{Gaebler16, Nichol17, Sheldon16, Casparis16, Barends14,
  Ballance16} single-qubit gates are accomplished with significantly
higher fidelity than two-qubit gates. A natural question is then, how
can we use the high-fidelity operations to diagnose other parts of the
quantum system? In this work, we propose such a procedure that uses a
single high-fidelity state initialization and a limited set of unitary
operations to diagnose other state preparations and measurement
operators.

The standard method to diagnose state preparations and measurements is
quantum tomography (QT). Quantum state tomography (QST) is a procedure
to estimate an unknown quantum state from experimental data. When the
measurements are informationally complete, the resulting data can be
used to estimate the corresponding density
matrix~\cite{Prugovecki77, Fuchs04}, and we say that the state is
``identifiable.'' Quantum detector tomography (QDT) is a procedure to
estimate an unknown quantum measurement operator~\cite{Luis99,
  Fuirasek01}. When the unknown measurement is applied to an
informationally complete set of already known quantum states, the
resulting data can be used to create an estimate of the measurement
operator, and the measurements are identifiable.

A drawback to standard QST and QDT is that they require well-known
measurements and state preparations respectively.  However, it is
difficult to produce such measurements and state preparations when
some processes have significantly lower fidelity than
others. Attempting standard QT in this case typically results in
estimates with systematic errors.

To combat systematic errors, we adapt standard QST and QDT to the
situation where a single state preparation and a limited set of
unitary operations (for example, single-qubit rotations) have
significantly higher fidelity than other processes and
measurements. For this situation, we develop a procedure to estimate
other states and the measurement operators simultaneously with an
alternating maximum likelihood estimation (MLE) algorithm. When the
measurements are not informationally complete the state is
non-identifiable. However, it is ``set identifiable,'' meaning that we
can specify a set of density matrices that are compatible with the
measurements. This set can be used to upper and lower bound
important quantities like fidelity of the state preparations or
expectation values of other observables. The estimates also provide
information about the quantum processes that produced the unknown
states.

Systems with a limited set of high-fidelity processes are common in
state-of-the-art quantum information experiments. Our original
motivation comes from trapped-ion experiments where single-qubit gates
have been demonstrated with high fidelity, but entangling gates have
lower fidelities. The procedure described here was implemented in
trapped-ion experiments described in Refs.~\cite{Lin16, Gaebler16}. We
return to the trapped-ion example throughout this paper to illustrate
our procedure. Trapped ions are measured by observing the presence
or absence fluorescence produced by ions in the ``bright'' or
``dark'' computational basis states.  A measurement datum is the
number of detected fluorescence photons.  To reduce computational
complexity, we coarse-grain the measurement outcomes with a strategy
that maximizes mutual information between the raw and coarse-grained
data.  Our strategy can be used for other qubit systems measured by
fluorescence or even systems that give continuous measurement
outcomes, like superconducting transmon qubits \cite{Reed10}.

QT with unknown states and measurements has been considered in
previous work~\cite{Mogilevtsev12,Mogilevtsev13}. Our procedure
differs from these works by using alternating MLE, which yields
estimates consistent with all the data collected.  Our procedure also
resembles other variants of QT, such as
self-consistent~\cite{Merkel13, Stark14} and gate-set tomography
(GST)~\cite{Blume-Kohout13}. These methods treat the quantum system as
a ``black-box'' about which nothing (or almost nothing) is known.
Notably, in GST one has access to a collection of unknown quantum
processes or ``gates,'' and one creates a series of experiments,
applying the gates in different orders. By performing the proper
experiments, one can find a full estimate of all the gates
simultaneously (up to an unobservable
gauge)~\cite{Blume-Kohout13}. The method has been successful in
experiments for single qubit systems~\cite{Medford13, Dehollain16,
  Blume-Kohout17}. In GST, the goal is to learn everything about the
system. In our procedure, we assume more about the system and ask less
about the outcome, thereby requiring fewer experiments. While this
requires stronger pre-experiment knowledge about the system than is
available in GST, we believe that it is well suited to a situation
that occurs in many state-of-the-art experiments and complements these
other proposals.

We begin by defining and describing the standard methods for QST and
QDT in Sec.~\ref{sec:General_model}. We also discuss and physically
motivate some additional assumptions about the measurement that
greatly simplify our procedure.  Then, in Sec.~\ref{sec:experiments},
we introduce the experiments required for our procedure and discuss
the application to the example trapped-ion system. Next, we discuss
our numerical technique to extract estimates of the states and
measurements from the experimental data in Sec.~\ref{sec:alg}. In
Sec.~\ref{sec:expectation_values}, we show how to upper- and
lower-bound important expectation values if the states are only
set-identifiable. We then discuss how to estimate the uncertainties
Sec.~\ref{sec:uncertainties}. We summarize our procedure and discuss
future directions in Sec.~\ref{sec:conclusions}. We also include three
appendices that describe our software implementation
(App.~\ref{app:software}), a method for coarse-graining measurement
outcomes (App.~\ref{app:binning}), and the stopping criteria for
iterations of the likelihood maximization (App.~\ref{app:stopping}).

\section{General model} \label{sec:General_model} Our procedure is
rooted in QST and QDT, so we begin with a brief description of the
standard version of each. We focus on quantum systems that are
described with finite, $d$-dimensional Hilbert spaces. QST is a
procedure to estimate the unknown $d \times d$ density matrix $\rho$,
that describes the state of a quantum system. To estimate $\rho$, we
prepare many identical copies of the quantum state and apply a known
quantum measurement to each.  In the following, a sequence of
identical state preparations and measurements is called an
``experiment'', while a particular state preparation and measurement
in an experiment is referred to as a ``trial.''  A quantum measurement
is associated with the measurement operators $F_{b}$ of a POVM
$\{ F_b \}_{b}$, where $F_b \geq 0$ and $\sum_b F_b = \mathds{1}$.
The probability of outcome $b$ in a trial is given by the Born rule
$p_b = \Tr{(F_b \rho)}$. In some instances, several different POVMs
are used for QST; in these cases we perform separate experiments for
each POVM. The QST formalism assumes that uncertainty about the
measurement operators of the POVMs is negligible. If the measurement
operators from all experiments span the bounded operators on the
Hilbert space, we call the set of POVMs ``informationally complete''
(IC).  If the state's probabilities for the outcomes of a set of IC
POVMs with specified measurement operators are exactly known, one has
all the information necessary to exactly reconstruct the unknown
density matrix~\cite{Fuchs04, Prugovecki77}.  In an experimental
implementation, the probabilities are never exactly known, because one
only runs a finite number of trials. One therefore numerically
estimates the quantum state from the measured frequencies of the
outcomes via techniques such as MLE~\cite{Hradil04, Rehacek07}.

In QDT, the goal is to estimate the measurement operators associated
with an unknown quantum measurement.  To estimate the measurement
operators, we prepare many identical copies of members of a family of
known quantum states $\{ \rho_j \}_{j}$. We then perform experiments
where we prepare many copies of one member of this family and apply
the unknown quantum measurements. For a given trial in one of these
experiments the probability of each outcome is given by the Born rule
$p_{j,b} = \Tr{(F_b \rho_j)}$.  In the QDT formalism, we assume the
uncertainty about the states $\{ \rho_j \}_{j}$ is negligible. We
define an IC set of states for QDT as a set that spans the bounded
operators on the Hilbert space, which is analogous to IC POVMs in
QST. The probabilities from each outcome along with the exact
description of an IC set of states allow for unique reconstruction of
each measurement operator. As with QST, in practice we cannot
determine the probability of each outcome due to a finite number of
copies of the unknown state. Thus, the measurement operators are
numerically estimated from the frequencies of the outcomes via
techniques such as MLE~\cite{Fuirasek01}.

In this work, we use aspects of QST and QDT to create a hybrid
procedure that estimates both states and detectors in a single
maximization of the likelihood given the entire data set.  Instead,
one could consider first performing QDT tomography using $\{ \rho_j
\}_{j}$.  Once the detectors have been calibrated, they could be
used for QST.  Compared to this separated strategy, our procedure
achieves lower uncertainty because it includes all data when
estimating both the measurement operators and unknown states.  (The
separated strategy would not include data from unknown states in
QDT.)  We obtain the global maximum of the full likelihood rather
than maximizing two separate likelihood functions, which may not be
maximized at the same location as the full likelihood function.
Furthermore, when using the separated strategy one must somehow
propagate uncertainty in the measurement operators into the QST,
which uses those measurement operators, but we do not know of a
robust method for that uncertainty propagation.

We assume that one has access to a limited set of
high-fidelity unitary operations. We designate such a set of unitary
operators with indices as
$\boldsymbol{\mathcal{U}} = \{ \mathcal{U}_i |i=1,\dots, r\}$, where
$\mathcal{U}_i [\rho] = U_i \rho U_i^{\dagger}$ and $U_0$ is the
identity, $U_0 = \mathds{1}$. We additionally assume that we can
reliably prepare a known state $\rho_0$.  Then, by applying the
high-fidelity processes to this state, we generate a family of known
states
$\boldsymbol{\rho} = \{ \rho_i = \mathcal{U}_i [\rho_0] \}_{i}$. There
could be repetitions in this family. In addition to these known
states, other processes generate a family of unknown states
$\boldsymbol{\sigma} = \{ \sigma_j|j=1,\ldots,s\}$ to be
characterized. All states are read-out by an unknown quantum
measurement described by the POVM
$\boldsymbol{F}=\{F_b|b=1,\dots, M \}$. To measure in different bases,
we apply one of the high-fidelity processes prior to the measurement,
thereby creating the measurement operators
$F_{i,b} = \mathcal{U}_i^{\dagger}[F_b] = U_i^{\dagger} F_b U_i$,
where $F_{0,b} = F_b$ by definition of $U_0$.

\begin{figure}
  \centering\includegraphics[width=3.4in]{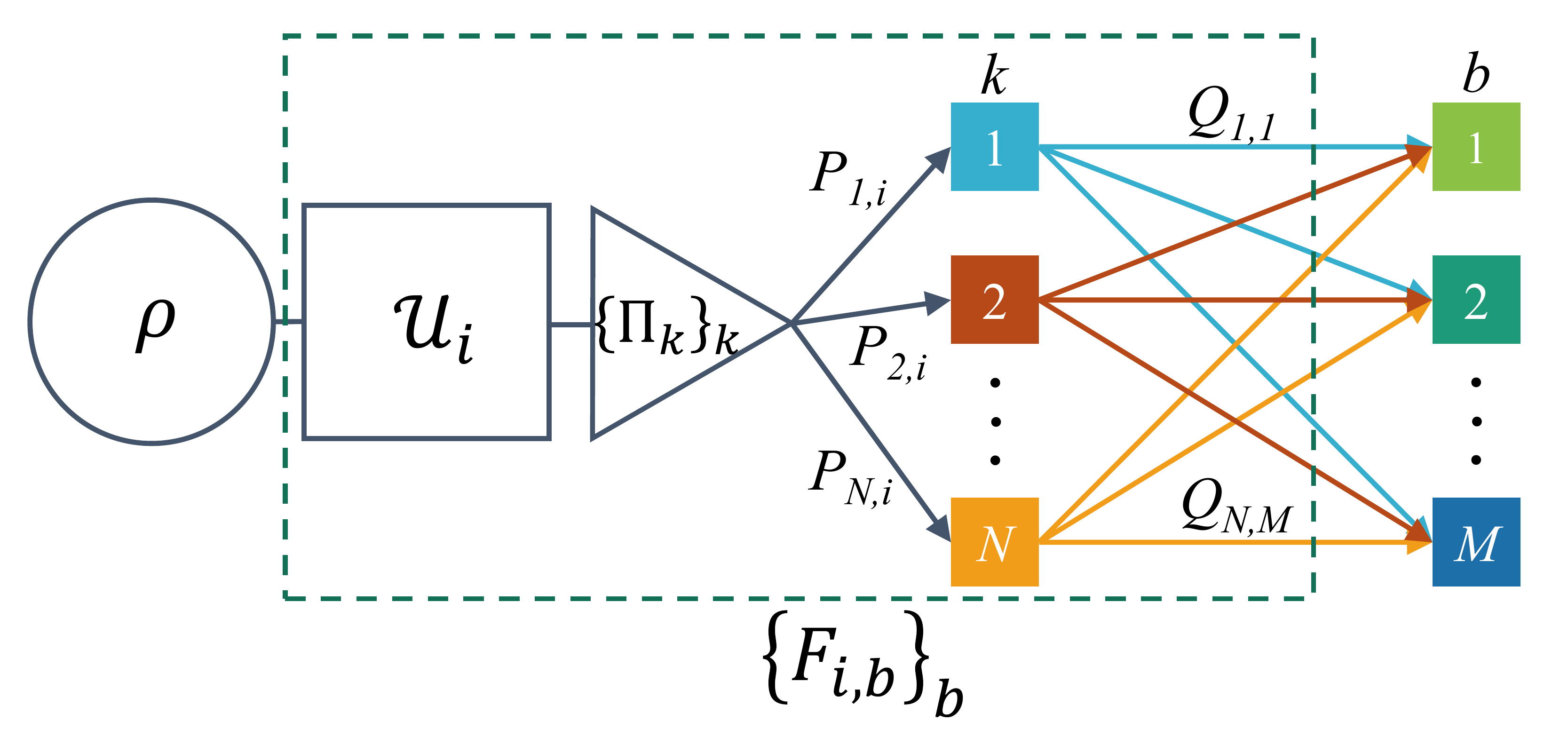}
  \caption{\label{fig:measurement_schematic} Schematic describing the
    measurement model for a single engineered measurement
    $\{F_{i,b}\}_{b}$. A density matrix $\rho$ is rotated by the
    unitary operator $U_i$, which determines the basis in which $\rho$
    is measured.  In the measurement, first the underlying POVM
    $\{ \Pi_{k} \}_{k}$ is applied, and result $k$ is obtained with
    probability $P_{k,i}$. Then, a Markov process acts to give random
    outcome $b$ with probability $Q_{k,b}$.  In the experiment we only
    perceive the outcome $b$.  Ellipses and dotted arrows show that
    there may be an arbitrary finite number of underlying measurement
    operators $\Pi_{k}$ and values of $b$.}
\end{figure}

To significantly simplify our procedure, we add an additional
assumption that the measurement operators $F_b$ are unknown mixtures
of the operators of an underlying POVM
$\boldsymbol{\Pi} = \{ \Pi_k |k = 1,\dots,N\}$, whose members are well
known.  To avoid numerical instabilities, we assume that no two
measurement operators in $\boldsymbol{\Pi}$ are equal. The resulting
measurement is depicted in Fig.~\ref{fig:measurement_schematic}.  We
model the measurement process by the application of the underlying
POVM $\boldsymbol{\Pi}$ yielding hidden outcome $k$, followed by an
unknown Markov process, after which outcome $b$ is observed. The
Markov process is described by the transition matrix $Q_{k,b}$.  (If
the outcome $b$ is a continuous variable, our procedure requires
discretization.) This situation is common in many experiments, where
physics constrains the underlying quantum measurement process, but
subsequent incoherent effects make identification of the measurement
operators corresponding to outcomes difficult. Previous techniques
have further constrained the model by assuming a form for the Markov
process. For example, in some ion trap experiments, the observed
outcomes are assumed to be a mixture of Poissonians given the hidden
outcomes~\cite{Itano93, Acton06}.  By not constraining the Markov
process by a model, we avoid systematic errors from model mismatch.

The measurement operator $F_b$ can now be expressed in terms of the
Markov process and the underlying measurement operators as
\begin{equation}\label{eq:singlemeasurement}
  F_b = \sum_k Q_{k,b} \Pi_{k}.
\end{equation}
The probability of observing $b$ given state $\rho$ is
$p_b = \Tr{(F_b \rho)}$. The POVM $\boldsymbol{F}$ is referred to as
the ``bare POVM.'' As mentioned above, we also apply the high-fidelity
processes $\boldsymbol{\mathcal{U}}$ prior to the measurement to
create ``engineered POVMs'' whose members are
\begin{equation}\label{eq:multimeasurement}
  F_{i,b} = \sum_k Q_{k,b} \Pi_{i,k},
\end{equation}
where $\Pi_{i,k} = \mathcal{U}_i^{\dagger}[ \Pi_k ]$. Because we
designated $U_0 = \mathds{1}$, $\{ F_{i,b} \}_{i,b}$ is the complete
family of measurement operators, bare and engineered.
 
In the theory described below, the underlying POVM $\boldsymbol{\Pi}$
may be any POVM, but we specifically discuss POVMs whose measurement
operators are orthogonal subspace projectors. This further assumption
was originally motivated by trapped-ion experiments where the
underlying POVM consists of projectors onto orthogonal internal states
of each ion. This model is a useful description because the
quantization axis and measurements are aligned with high accuracy
(further details are given in the next section).  The situation also
occurs in other experiments with similarly high accuracy alignment. If
the assumption does not hold naturally, it can be easily enforced if
the operations take place in a rotating frame. Then the measurements
can be decohered by randomizing the phase (or time) between the
operations and measurements for each trial. Then, averaged over many
trials any coherence will be lost and the underlying measurement
operators are orthogonal subspace projectors.

\section{Experimental measurement procedure} \label{sec:experiments}
\begin{figure*}
  \centering
  \includegraphics[width=\textwidth]{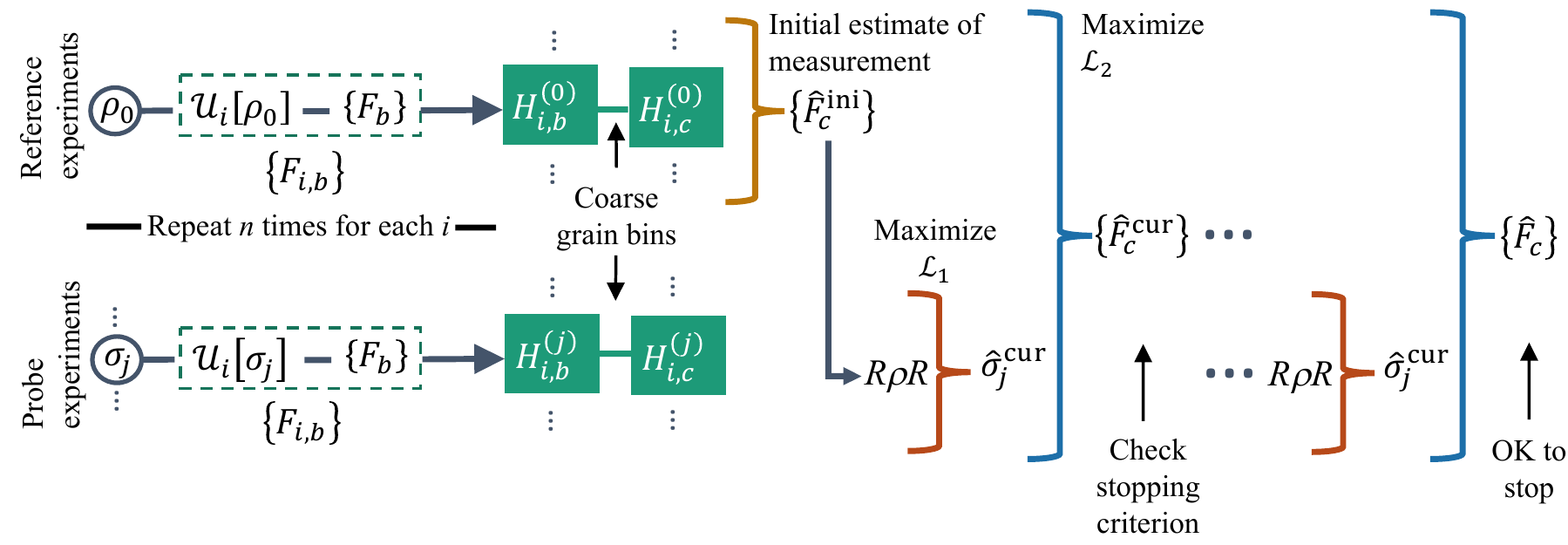}
  \caption{\label{fig:inference_chain} Schematic showing measurement
    and estimation procedure.  Reference experiments begin with the
    preparation of the known state $\rho_0$, and probe experiments
    begin with unknown states from
    $\boldsymbol{\sigma} = \{\sigma_j |j=1, \dots, s\}$.  During each
    experiment the states are transformed by a process in the set
    $\boldsymbol{\mathcal{U}} = \{\mathcal{U}_i | i = 0 \dots r\}$ and
    then measured with the POVM
    $\boldsymbol{F} = \{F_b| b=1,\dots,M\}$, giving outcome $b$.  For
    simplicity, we assume every known process acts on each unknown
    state, but this is not necessary. After $n$ trials of each
    experiment, histograms $\{H_{b,i}^{(j)} \}_{b,i}$ for each state
    are recorded.  These histograms are then coarse-grained to produce
    new histograms $\{H_{c,i}^{(j)} \}_{c,i}$ based on a training data
    set composed of 10~\% of the trials randomly selected without
    replacement from each reference experiment. From the reference
    experiments' data, we calculate an initial estimate for the POVM
    $\hat{\boldsymbol{ F}}^{\text{ini}}$.  We divide the likelihood
    maximization into two concave subproblems: maximization of
    $\mathcal{L}_1$ with respect to $\boldsymbol{\sigma}$ and
    maximization of $\mathcal{L}_2$ with respect to
    $\boldsymbol{F}$. We alternate between using the $R \rho R$
    algorithm to maximize $\mathcal{L}_1$ and a standard nonlinear
    optimizer to maximize $\mathcal{L}_2$.  At each iteration we find
    maximizing parameters $\hat{\boldsymbol{\sigma}}^{\text{cur}}$ and
    $\hat{\boldsymbol{F}}^{\text{cur}}$, and we check a stopping
    criterion.  When the stopping criterion signals that we can stop
    iterations, we have the numerical maximum likelihood estimates
    $\hat{\boldsymbol{\sigma}}$ and $\hat{\boldsymbol{F}}$. }
\end{figure*}

Our procedure consists of two experimental steps followed by numerical
processing of the data. In this section, we describe the experimental
steps and introduce an example application to trapped-ion systems to
illuminate the discussion.  A schematic for the measurement and
numerical estimation procedure is in Fig.~\ref{fig:inference_chain}.

The first experimental step is related to standard QDT of the bare
POVM. We call the experiments done during this
step the ``reference'' experiments since they will be used to
initialize the algorithm described in the next section.
There is one reference experiment for each known pure state $\rho_{i}$ in $\boldsymbol{\rho}$.
The $i$th reference experiment, indexed $i$, consists of $n$ trials. 
In each trial, we prepare $\rho_i$ and apply
the bare POVM. The probability for outcome $b$ in a given trial is
\begin{equation} \label{exp1} p_{i,b}^{(0)} = \Tr{( F_b \rho_i )} =
  \sum_k Q_{k,b} \textrm{Tr}(\Pi_k \rho_i).
\end{equation}
The measurement outcomes are sampled from the distribution given by
Eq.~\eqref{exp1}. We assume the outcomes of the trials are independent
and identically distributed. The outcomes of all trials from all
reference experiments are collected into a matrix $H^{(0)}$ such that
row $i$ is the relative frequency histogram for experiment $i$.
Specifically, the matrix element $H^{(0)}_{i,b}$ is the number of
times outcome $b$ is observed for experiment $i$ divided by $n$. We
call this matrix the ``histogram matrix'' in the following.

The second experimental step is similar to QST, where the goal is to
extract information about the unknown quantum states
$\boldsymbol{\sigma}$. We call the experiments done during this step
the ``probing experiments,'' because they are designed to probe the
unknown states.  There is one probing experiment for each engineered
POVM and unknown state. The probing experiment, indexed by $i,j$,
consists of $n$ trials, where in each trial, we prepare $\sigma_{j}$
and apply the engineered POVM indexed by $i$.  The probability of
getting outcome $b$ in a given trial is
\begin{equation} \label{exp2} p_{i,b}^{(j)} = \textrm{Tr}(
  \mathcal{U}_i^{\dagger}[F_b] \sigma_j) = \sum_k Q_{k,b}
  \textrm{Tr}(\Pi_{i,k} \sigma_j).
\end{equation} The outcomes of the trials from the probing experiments
on state $\sigma_j$ are collected into the histogram matrix $H^{(j)}$.

In general, the probability of outcome $b$ in a trial of a reference
or probing experiment has the probability distribution,
\begin{equation} \label{total_prob} p_{i,b}^{(j)} = \textrm{Tr}(
  \mathcal{U}_i^{\dagger}[F_b] \tau_{j}) = \sum_k Q_{k,b}
  \textrm{Tr}(\Pi_{i,k} \tau_j),
\end{equation}
where
$\boldsymbol{\tau} = (\tau_j)_{j=0}^{s} = (\rho_0, \sigma_1, \dots,
\sigma_s )$.  For $j = 0$, Eq.~\eqref{total_prob} reduces to
Eq.~\eqref{exp1} and for $j >0$ it reduces to Eq.~\eqref{exp2}.

In this discussion we have made a few simplifying assumptions about
the measurement record, such as that every experiment contains the
same number of trials and that every state $\tau_j$ is subjected to
the same set of measurements. These assumptions are not necessary for
our procedure and are made here only to simplify the mathematical
notation.

As an example of the complete procedure we consider the task of
diagnosing the internal state of two trapped ions, similar to
Ref.~\cite{Gaebler16}. Each ion is treated as a single qubit with
computational basis $\ket{\uparrow}$, called the ``bright state,'' and
$\ket{\downarrow}$, called the ``dark state.'' Optical pumping is used
to initialize each ion in the bright state,
$\rho_{0} = \ket{\uparrow \uparrow} \bra{ \uparrow \uparrow }$, with
high fidelity.

As in Ref.~\cite{Gaebler16}, we consider the high-fidelity processes
to be single-qubit rotations applied to both ions. A single-qubit
rotation is defined by
\begin{equation}
  U(\theta, \phi) = \textrm{exp}\left[-i \tfrac{\theta}{2} (\sigma_x \cos\phi + \sigma_y \sin \phi) \right],
\end{equation}
where $\sigma_x$ and $\sigma_y$ are Pauli operators. Collective
rotations $U(\theta, \phi)^{\otimes 2}$ can be accomplished with,
e.g., microwaves applied uniformly across the ion trap. For the
example discussed below, we choose the subset
$ \boldsymbol{U} = \{U(0,0)^{\otimes 2}, U(\tfrac{\pi}{2},0)^{\otimes
  2}, U(\pi,0)^{\otimes 2}, U(\tfrac{\pi}{2}, \tfrac{\pi}
{2})^{\otimes 2} \}$ as the set of high-fidelity processes.  Since
two-qubit entangling operations have significantly lower fidelities, a
natural task is to diagnose entangled states created by these
operations.

Measurement of the ions is accomplished by stimulating a cycling
transition between the bright state and another internal state
(outside of the computational basis). This transition is well aligned
with the quantization axis and highly detuned from the dark state,
which implies observing fluorescence is a projective measurement of an
ion in the bright state. The ions are close together in the trap, so
the measurement cannot distinguish which ion is fluorescing. The
corresponding POVM consists of subspace projectors, $\boldsymbol{\Pi}
= \{\Pi_0 = \ket{\downarrow \downarrow}\bra{\downarrow \downarrow},
\Pi_1 = \ket{\downarrow \uparrow}\bra{\downarrow \uparrow} +
\ket{\uparrow \downarrow}\bra{\uparrow\downarrow}, \Pi_2 =
\ket{\uparrow \uparrow}\bra{\uparrow\uparrow}\}$. However, we cannot
observe the outcome of this POVM directly in any experiment since the
number of photons in the fluorescence signal is distributed according
to counting statistics as well as other detection errors such as dark
counts, re-pumping to the dark state, or detector inefficiency. These
effects act as the Markov process described above, so the observed
outcomes are associated with the POVM, $\boldsymbol{F}$.

\begin{figure*}
  \centering
  \includegraphics[width=\textwidth]{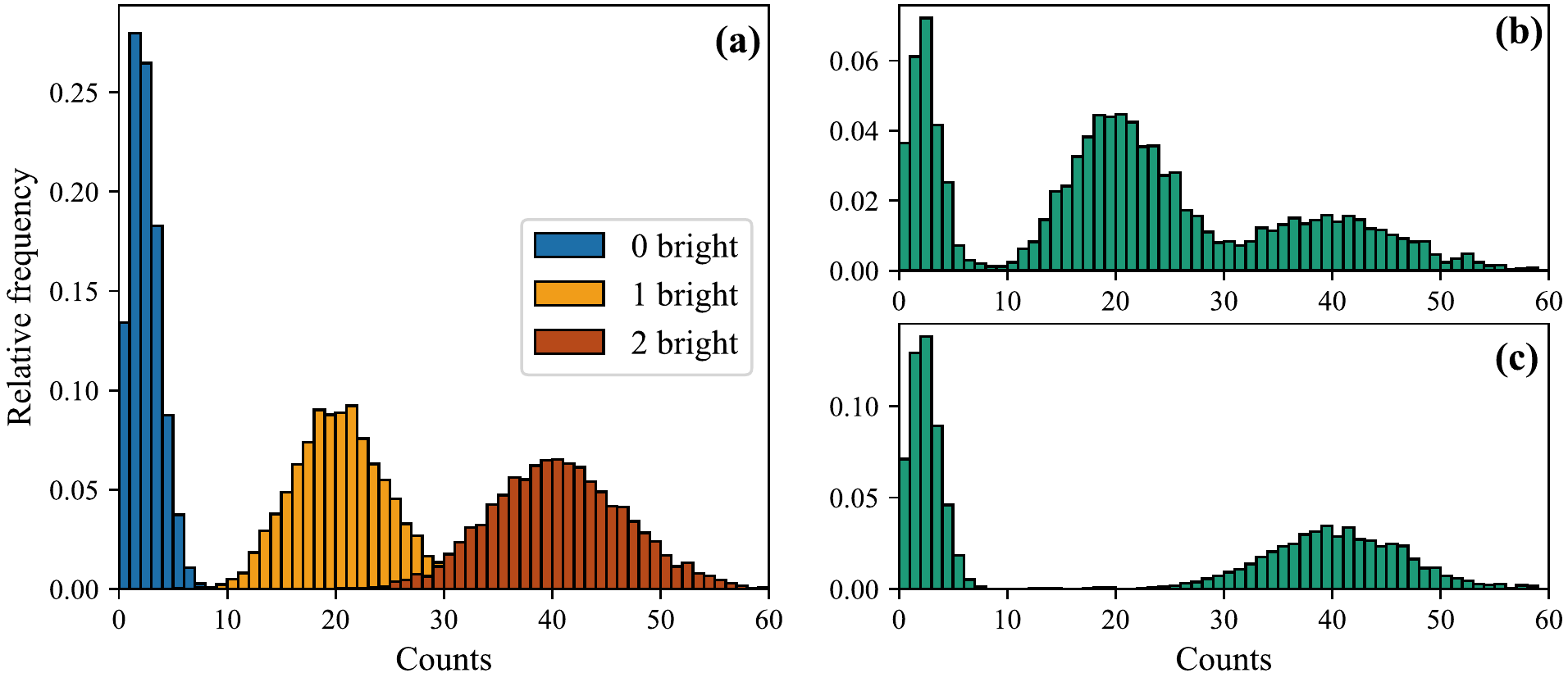}
  \caption{\label{fig:hists} Histograms from simulated measurements of
    two ions. An individual ion fluoresces if it is in the
    $\ket{\uparrow}$ (bright) state. We model the distributions
    of photon counts as Poissonians.  In a real experiment the
    distributions differ from Poissonians due to processes such as
    repumping from the dark to the bright state. For two ions, there
    are three possible count distributions corresponding to the total
    number of ions in the bright state: zero ions bright (mean 2, due
    to dark counts), one ion bright (mean 20), and two ions bright
    (mean 40). For one ion bright, the photon detector cannot
    determine which of the two ions is fluorescing. (a) Relative
    frequency histograms for both ions in the dark state (blue), one
    ion on the dark state (orange), and both ions in the bright state
    (red). (b) Relative frequency histogram for the reference
    experiment with $U_1 = U(\tfrac{\pi}{2}, 0)^{\otimes 2}$. (c)
    Relative frequency histogram for the probing experiment with $U_3
    = U(\tfrac{\pi}{2},\tfrac{\pi}{2})^{\otimes 2}$.}
\end{figure*}

Simulated histograms of the above states are shown in
Fig.~\ref{fig:hists}. With the processes considered, we cannot create
an IC set of engineered POVMs because the POVM $\boldsymbol{\Pi} $
(and therefore also $\boldsymbol{F}$) cannot distinguish the two
single-ion-bright states and all of the rotations in
$\boldsymbol{\mathcal{U}}$ act equally on both qubits. However, the
measurements are sufficient to identify a Bell-state fidelity, as is
discussed in Sec.~\ref{sec:expectation_values}.

\section{Maximum likelihood estimation technique} \label{sec:alg} 

We now present a maximum likelihood estimation (MLE) technique that
produces numerical estimates of the measurement operators and the
unknown states. Like all MLE techniques, if the true operators and
states are not on the boundary, the technique is optimal in that its
variance asymptotically approaches the Cram\'{e}r-Rao lower
bound~\cite{Shao98}. Our algorithm starts by coarse graining the
outcome space $b$, which condenses the data to speed up the numerical
processing. We then create an initial estimate of the transition
matrix $Q$. Finally, we search for the estimates of $Q$ and
$\boldsymbol{\sigma}$ that maximize the total log-likelihood function
of all the data. This is accomplished by an alternating approach.
While this is not proven to yield the maximally likely operators and
states in all instances, we have found no counterexamples and believe
that the maximum likelihood solution is typically found.
  
The first step is to coarse grain the outcome space.  In many cases,
the set of possible values of the outcomes $b$ is large, that is $b =
1,\dots,M$ where $M \gg 1$.  The collected histograms and transition
matrix are then also large, which is a main contributor to the
computational cost of likelihood maximization.  Coarse graining the
outcome space shrinks the histograms and transition matrix, thereby
reducing the complexity. The coarse graining is accomplished by
creating a pre-specified number $G < M$ of contiguous bins on the set
of possible outcomes.  Each bin is defined by its edges in the outcome
space.  We index the bins, which correspond to new coarse-grained
outcomes, by $c = 1,\dots, G$.  Each outcome $b$ is then reassigned to
a coarse-grained outcome $c$ based on the bin edges.  The bin edges
are constructed according to a heuristic that minimizes the
information loss between the original outcomes and the coarse grained
outcomes (details are in Appendix~\ref{app:binning}).  We randomly
select without replacement 10~\% of the trials from the reference
experiments as a training set with which we choose the bin edges.  The
training set is then excluded from further analysis.  (To simplify
notation, in the results presented below, we simulated reference
experiments with $(10/9)n$ trials so that the number of trials for
reference and probing experiments is equal after coarse graining,
although the technique can be used with unequal numbers of trials.)
The coarse-grained outcomes then determine a new set of histogram
matrices $H_{i,c}^{(j)}$, measurement operators $F_c$, and transition
matrix $Q_{k,c}$.  For the remainder of the paper we discuss the
coarse-grained version of each.

The next step is to use the reference histogram matrix $H^{(0)}$ to
derive an initial estimate of the transition matrix $Q$, which we
later use to initialize the likelihood maximization.  In the limit of
infinite trials $H^{(0)}$ is equal to the probability distribution
from Eq.~\eqref{exp1} but with a finite number of trials
$H^{(0)}_{i,c} \approx \sum_k Q_{k,c} \textrm{Tr}(\Pi_k \rho_i)$. We
can re-express this relation in matrix form, $H^{(0)} \approx P Q$,
where $Q$ has elements $Q_{k,c}$, and $P$ has elements
$P_{i,k} = \textrm{Tr}(\Pi_k \rho_i)$. We refer to the elements
$P_{i,k}$ as ``populations,'' since they are the probabilities that
state $i$ is in subspace $k$. The population matrix $P$ is known {\it
  a priori}, because the states are prepared with the high-fidelity
processes. Therefore, we can derive an initial estimate for $Q$, which
describes the bare measurement, by applying the left pseudo-inverse of
$P$,
\begin{equation} \label{q_est} \hat{Q}^{\text{ini}} = \left( P^{\top}
    P\right)^{-1} P^{\top} H^{(0)} .
\end{equation}
The left pseudo-inverse exists when $\textrm{rank}(P) = N$, that is,
the number of subspace projections.  This requires that the states
span the subspace that is spanned by the underlying POVM.  We only
consider situations where this condition is met. The initial estimate
of the transition matrix is also used to calculate an initial estimate
of the observed measurement operators,
$\hat{F}_c^{\text{ini}} = \sum_k \hat{Q}^{\textrm{ini}}_{k,c} \Pi_k $.
    
The final step is to determine estimates of the states and
measurement operators that maximize the total likelihood function
$L$. To derive $L$, we first need the probability of observing a given
histogram matrix $H^{(j)}$, which is the product of the probabilities
of observing each outcome from Eq.~\eqref{total_prob} and is given by
\begin{equation}
  \label{prob_eq}
  \Prob(H^{(j)} | \boldsymbol{\tau},
  \boldsymbol{F},\boldsymbol{\mathcal{U}}) = \prod_{i,c}
  \Tr{(\mathcal{U}^{\dagger}_i[F_{c}] \tau_{j})}^{H^{(j)}_{i,c}}.
\end{equation}
The probability of obtaining the histogram matrix from all of the
experiments given the unknown and the known parameters is then
\begin{equation}
  \Prob(\boldsymbol{H} | \boldsymbol{\tau}, \boldsymbol{F}, \boldsymbol{\mathcal{U}}) =  \prod_{j,i,c} \Tr{(\mathcal{U}
    ^{\dagger}_i[F_{c}] \tau_j)}^{H^{(j)}_{i,c}},
\end{equation}
where $\boldsymbol{H} = \{H^{(j)} \}_j$. This probability is the total
likelihood function $L$. When we maximize $L$, we vary the unknown
parameters $\boldsymbol{\sigma}$ and $\boldsymbol{F}$, which are the
parameters of the statistical model, and keep the known parameters
$\rho_0$, $\boldsymbol{\mathcal{U}}$ and the histogram matrix
$\boldsymbol{H}$ fixed. To emphasize the distinction between varying
and fixed parameters, we write the likelihood function as
$L(\boldsymbol{\sigma}, \boldsymbol{F} |\boldsymbol{H},
\boldsymbol{\mathcal{U}},\rho_0) = \Prob(\boldsymbol{H}|
\boldsymbol{\tau}, \boldsymbol{F}, \boldsymbol{\mathcal{U}})$.  The
estimates for the states and measurement operators that maximize $L$
also maximize the log-likelihood function $\lik = \ln(L)$. As is
standard practice, we maximize $\lik$ instead of $L$ because $\lik$
has convenient concavity properties, and it avoids numerical issues
with extremely small values. The log-likelihood is
\begin{equation}
  \label{eq:loglike}
  \lik(\boldsymbol{\sigma}, \boldsymbol{F}|\boldsymbol{H}, \boldsymbol{\mathcal{U}},\rho_0) = \sum_{i,j,c} H_{i,c}^{(j)} \ln 
  \Tr{(\mathcal{U}^{\dagger}_i[F_c] \sigma_j)}.
\end{equation}
In this case, the log-likelihood is not a concave function, which makes
it difficult to determine the global maximum value. However, $\lik$ is
separately concave in $\boldsymbol{\sigma}$ and $\boldsymbol{F}$. This
can be confirmed by computing the log-likelihood's second derivatives
with respect to $\boldsymbol{\sigma}$ and $\boldsymbol{F}$ and noting
that they are negative semidefinite.

We present an iterative technique to find the maximum of $\lik$.
Since the log-likelihood is separately concave in
$\boldsymbol{\sigma}$ and $\boldsymbol{F}$, optimizing over one while
holding the other fixed is a concave optimization problem whose local
maxima are global maxima. So to maximize $\lik$ over both $\sigma$ and
$\boldsymbol{F}$ jointly, we alternate between two subproblems: (1)
concave optimization over $\boldsymbol{\sigma}$ with $\boldsymbol{F}$
fixed and (2) concave optimization over $\boldsymbol{F}$ with
$\boldsymbol{\sigma}$ fixed (details are given below).  Here, we
describe the subproblem optimizations in the case of a single unknown
state, $ \boldsymbol{\sigma} = \{ \sigma \}$ (we drop the boldface
notation for the unknown states since the family has a single member),
which requires a single set of probing experiments with histogram
matrix $H^{(1)}$.

The first subproblem is maximization of the log-likelihood with respect
to the unknown density matrix $\sigma$ keeping the measurement
operators fixed at their current estimate
$\boldsymbol{\hat{F}}^{\text{cur}}$. For the initial step, we fix the
measurement as $\boldsymbol{\hat{F}}^{\text{cur}} =
\boldsymbol{\hat{F}}^{\text{ini}}$.  For this subproblem the objective
function is
\begin{equation} \label{loglike_1}
  \lik_1(\sigma|H^{(1)},\boldsymbol{\mathcal{U}},\boldsymbol{\hat{F}}^{\text{cur}})
  = \sum_{c,i} H^{(1)}_{c,i} \ln
  \Tr{(\mathcal{U}^{\dagger}_i[\hat{F}_c^{\text{cur}}] \sigma)}.
\end{equation}
Note that $\lik_1$ only uses the histogram matrix $H^{(1)}$ from the
probing experiments because the reference experiments are independent
of $\sigma$.  We numerically search for the density matrix $\sigma$
that maximizes $\lik_{1}$ by solving
\begin{equation}
  \begin{aligned}
    & \underset{\sigma}{\text{maximize:}}
    & & \lik_1(\sigma|H^{(1)},\boldsymbol{\mathcal{U}},\boldsymbol{\hat{F}}^{\text{cur}}), \\
    & \text{subject to:}
    & & \Tr{\sigma} = 1, \\
    &&& \sigma \succeq 0,
  \end{aligned}
\end{equation}
where $\sigma \succeq 0$ means that $\sigma$ is a positive
semidefinite matrix.  To accomplish this, we use the $R \rho R$
algorithm~\cite{Rehacek07,Goncalves14}, initialized with
$\sigma^{(1)} = \hat{\sigma}^{\textrm{cur}}$
($\hat{\sigma}^{\textrm{cur}} = \mathds{1}/d$ for the first
iteration).  At the $k$th iteration of the algorithm, the state
$\sigma^{(k)}$ is updated to
$\sigma^{(k+1)}=\mathcal{N}(R(\sigma^{(k)})\sigma^{(k)}R(\sigma^{(k)}))$,
where $\mathcal{N}$ indicates normalization and $R$ is
\begin{equation}
  R(\sigma)  = \sum_{i,c} \frac{H^{(1)}_{i,c} \hat{F}_{i,c}^{\text{cur}}}{ \Tr{(\sigma \hat{F}_{i,c}^{\text{cur}})}}.
\end{equation}
This ensures that at each iteration the estimate is physical and the
likelihood is nondecreasing \cite{Rehacek07}.  The algorithm is run
until the stopping condition derived in Ref.~\cite{Glancy12} is
met. For multiple unknown states, $ \lik_1$ is the sum of the
log-likelihoods for each state $\lik_1 = \sum_j \lik_1(\sigma_j),$
where $\lik_1(\sigma_j)$ has the form of Eq.~\eqref{loglike_1}. To
maximize $\lik_1$ in this case, we calculate $R$ for each $\sigma_j$
and iteratively update each $ \sigma_j$ individually.

The second subproblem is maximization of the log-likelihood with
respect to the measurement operators, $\boldsymbol{F}$, with $\sigma$
fixed at its current estimate $\hat{\sigma}^{\text{cur}}$ returned by
the most recent use of the $R \rho R$ algorithm. The objective
function for this subproblem is
\begin{equation}
  \lik_{2}(\boldsymbol{F}|\boldsymbol{H},\boldsymbol{\mathcal{U}},\boldsymbol{\hat{\tau}}^{\text{cur}}) = \sum_{c,i,j}  H^{(j)}
  _{c,i} \ln \Tr{(\mathcal{U}_i^\dagger[F_{c}] \hat{\tau}^{\text{cur}}_{j})},
\end{equation}
where
$\boldsymbol{\hat{\tau}}^{\text{cur}}=(\rho_{0},\hat{\sigma}^{\text{cur}})$. Because
the measurement operators are constrained according to
Eq. \eqref{eq:multimeasurement}, we re-write the objective function as
\begin{align}
  \lik_{2}(Q|\boldsymbol{H},\boldsymbol{\hat{P}}^{\text{cur}}) &=
  \sum_{c,i,j} H^{(j)}_{c,i} \ln\left(\sum_{k}Q_{k,c}\hat{P}
    ^{\text{cur}}_{k,i,j}\right),
\end{align}
where $\boldsymbol{\hat{P}}^{\text{cur}} = \{
\Tr{(\mathcal{U}_i^{\dagger}[\Pi_k] \hat{\tau}^{\text{cur}}_j)}
\}_{k,i,j}$.  We use a standard nonlinear multi-variable optimizer to
solve the following problem:
\begin{equation} \label{Q_program}
  \begin{aligned}
    & \underset{Q}{\text{maximize:}}
    & & \lik_{2}(Q|\boldsymbol{H}, \boldsymbol{\hat{P}}^{\text{cur}}), \\
    & \text{subject to:}
    & & \sum_c Q_{c,k} = 1 , \; \forall \, k, \\
    &&& 0 \leq Q_{c,k} \leq 1, \; \forall \, k,c.
  \end{aligned}
\end{equation}
The program is initialized with the current estimate of the transition
matrix $\hat{Q}^{\text{cur}}$ ($\hat{Q}^{\text{ini}}$ for the first
iteration) and runs until a pre-specified stopping tolerance is
reached. All histogram matrices are considered in the optimization,
which ensures that the estimated transition matrix $\hat{Q}$ is
consistent with both the reference and probing experiments described
in the previous section.

The assumption that the measurement operators are linear combinations
of an underlying POVM significantly simplifies the program in
Eq.~\eqref{Q_program}. Without this assumption, we would require
semi-definite constraints to ensure that the measurement is
physical. With the assumption, the linear constraints ensure that the
estimated transition matrix $ \hat{Q}$ is a probability distribution
and thus the corresponding observed measurement operators
$\boldsymbol{\hat{F}}$ are physical.

We alternate between these two subproblems until a gradient-based
stopping criterion is met (see Appendix~\ref{app:stopping}), yielding
estimates $\hat{\sigma}$ and $ \hat{\boldsymbol{F}}$. While there is
no guarantee that these estimates correspond to a global
maximum~\cite{Jochen07}, in numerical experiments we
find the algorithm converges to estimates that are close to the true
parameters.

There may be states other than $\hat{\sigma}$ that produce the same
maximum value of the likelihood function.  This occurs when the
high-fidelity processes do not generate an IC set of POVMs, that is,
$\{ \hat{F}_{i,c} \}_{i,c}$ does not span the space of bounded
operators.  This is the case in the trapped-ion example given at the
end of Sec.~\ref{sec:experiments}, where the POVM cannot determine
which ion is in the bright state.  In this case, $\hat{\sigma}$ is an
element of the ``set of MLE states'' where each element of the set
produces the same maximum value of the likelihood function.
Similarly, there may be transition matrices other than $\hat{Q}$ that
yield the same maximum value of the likelihood function. This occurs
when the family of known input states does not span the subspace of the
underlying POVM. However, as mentioned previously in this section, we
do not consider this case because it would cause other complications
with our initialization and would complicate the estimation of
expectation values described in the next section.

\section{Estimation of expectation
  values\label{sec:expectation_values}}

To gain information about the unknown states, we can estimate the
expectation value $\braket{O} = \Tr(O \sigma)$ of any observable
$O$. We can calculate the observable's expectation value according to
the point estimate from the likelihood maximization as
$ \hat{\braket{O}} = \Tr{(O \hat{\sigma})}$. However, if the set of
POVMs is not IC then it is possible that the set of MLE states
contains more than one element, and each element may have a different
expectation value with respect to $O$. This leads to a set of possible
expectation values for the unknown state, which are all equally
likely. Nevertheless, we can still learn about the unknown states by
determining bounds on the expectation value. This is accomplished by
searching for the elements in the MLE set that provide the largest and
smallest expectation values, which corresponds to the pair of
semidefinite programs (SDPs)
\begin{equation}
  \begin{aligned}
    & \underset{\rho}{\text{minimize:}}
    & & \pm \Tr{(O \rho)}, \\
    & \text{subject to:}
    & & \Tr{\rho} = 1,  \\
    &&&  \rho \succeq 0, \\
    &&& \Tr{\left(\hat{F}_{i,c}(\rho - \hat{\sigma}) \right)} = 0,
    \textrm{ for all $i,c$.}
  \end{aligned}
\end{equation}
The last constraint ensures that $\rho$ is in the set of MLE states by
constraining the expected probability of each outcome to be equal to
that of $\hat{\sigma}$. When this is the case, the log-likelihood of
any $\rho$ obeying the constraint is equal to the log-likelihood of
$\hat{\sigma}$.

As formulated above, the last line of the SDP contains
$|\{F_{i,c}\}_{i,c}| $ constraints, many of which may be redundant,
thereby causing increased computation time.  We reduce the number of
constraints by finding an orthogonal basis that spans
$\{ \hat{F}_{i,c} \}_{i,c}$. This is done by defining a size
$|\{F_{i,c}\}_{i,c}| \times d^2$ matrix $\boldsymbol{\mathcal{F}}$,
where each row is the measurement operator $\hat{F}_{i,c}$ written as
a $1 \times d^2$ vector.  We then calculate the singular value
decomposition,
$\boldsymbol{\mathcal{F}} = \boldsymbol{W}\boldsymbol{S}
\boldsymbol{V}^{\dagger}$.  The measurement operators
$\{ \hat{F}_{i,c} \}_{i,c}$ are in the span of the rows of
$\boldsymbol{V}^{\dagger}$.  To reduce the number of constraints in
the SDP, we create $\boldsymbol{V}'^{\dagger}$ by discarding the rows
of $\boldsymbol{V}^{\dagger}$ corresponding to the $\ell$ smallest
singular values, where the value of $\ell$ is equal to $d^2$ minus the
number of nonzero singular values of the related matrix formed by the
underlying measurement operators, $\{\Pi_{i,k}\}_{i,k}$.  Now, we
replace the last constraint in the above SDP with
$\Tr{[V'^\dagger_k(\rho - \hat{\sigma}) ]} = 0 $, where $V'^\dagger_k$
is the matrix formed from the $k$th row of
$\boldsymbol{V}'^\dagger$. This reduces the number of these
constraints to $d^2$ or fewer.

\begin{figure}
  \includegraphics[width=3.5in]{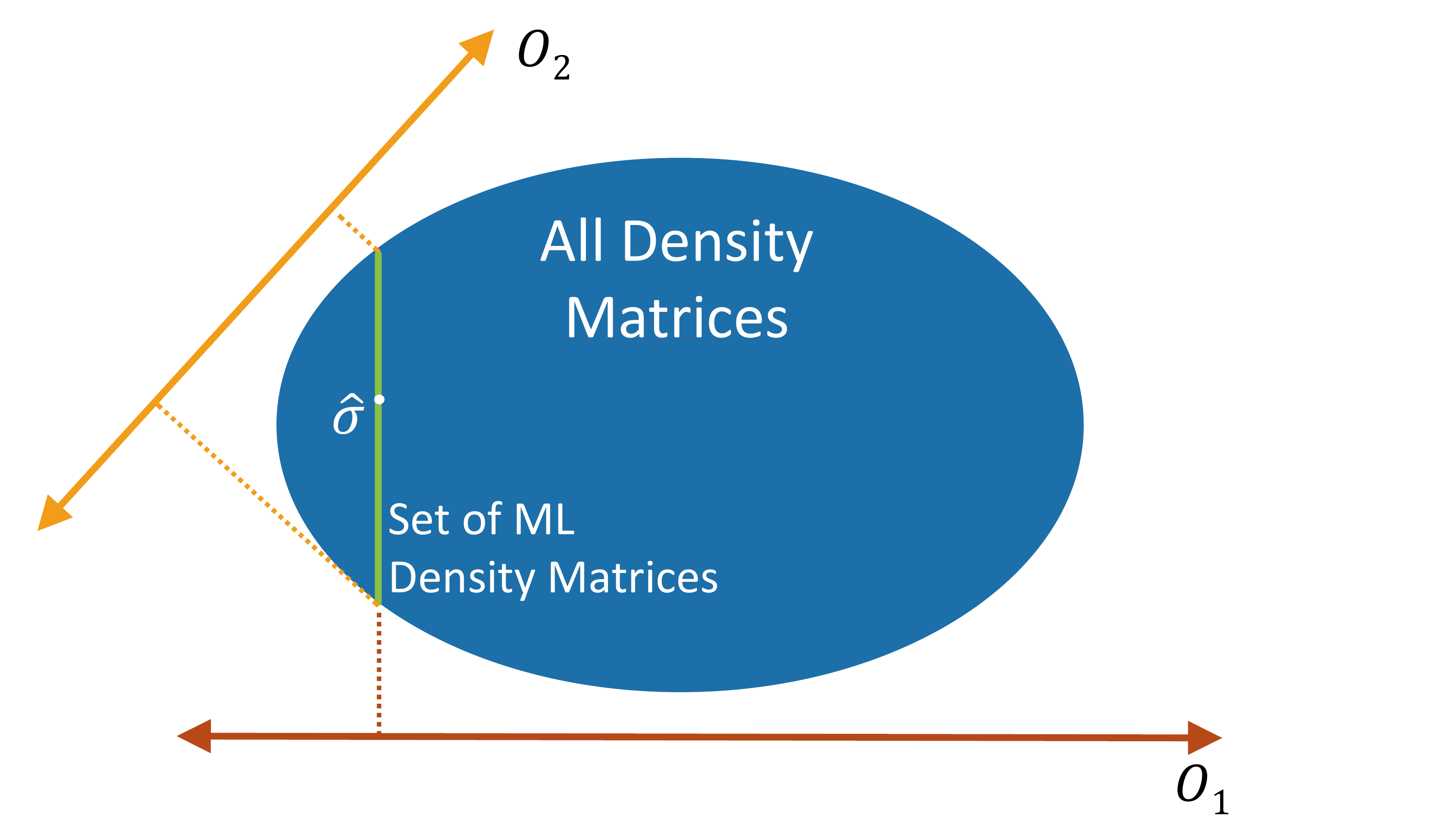}
  \caption{\label{fig:partialtomography} Conceptual schematic of
    bounding expectation values.  The likelihood maximization returns
    an estimate $\hat{\sigma}$ that is an element of the set of
    density matrices that maximize the likelihood (green line).
    Projecting from the set of maximum likelihood density matrices
    onto the line representing observable $O_1$ gives a unique
    expectation value. Because $O_1$ is in the span of the estimated
    measurement operators, each maximum likelihood density matrix has
    the same expectation value for $O_1$. Observable $O_2$ is only
    partially contained within the span of estimated measurement
    operators.  The SDP provides upper and lower bounds of its
    expectation value. }
\end{figure}

Observables can be divided into two types, identifiable and
set-identifiable, shown in Fig.~\ref{fig:partialtomography}.  First,
identifiable observables have the same expectation value for every
element in the set of MLE states. In this case the two SDPs return the
same expectation value. An identifiable observable $O_1$ is in the
span of the estimated measurement operators, namely there exist
coefficients $f_{i,c}$ such that
$O_1 = \sum_{i, c} f_{i,c} \hat{F}_{i,c}$.  The expectation values are
then proportional to the estimated outcome probabilities
\begin{equation}\label{eq:expectationdecomposed}
  \hat{\braket{O}} = \Tr{(O  \hat{\sigma})} = \sum_{i,c} f_{i,c} \Tr{(\hat{F}_{i,c}  \hat{\sigma})}.
\end{equation}
Since $\Tr{(\hat{F}_{i,c} \hat{\sigma})}$ has an associated empirical
estimate given by $H_{c,i}^{(1)}$, we can directly calculate
$\hat{\braket{O}}$. When $\{ F_{i,c} \}_{i,c}$ are IC, they span the
space of Hermitian operators, and every observable is identifiable. We
do not explicitly check if a decomposition of the form used in
Eq.~\eqref{eq:expectationdecomposed} exists apart from running the
SDPs.

Set-identifiable observables have a range of expectation values over
the MLE states. If $\{ F_{i,c} \}_{i,c}$ does not span all Hermitian
operators, then there necessarily exist observables that are
set-identifiable. The SDPs provide tight upper and lower bounds on the
ranges of the expectation values.

In the two-ion example, the measurements are not informationally
complete so there are both identifiable and set-identifiable
observables. An example of an identifiable observable is the Bell
state, $O_1 = \ket{ \Phi^+}\bra{ \Phi^+}$, where
$ \ket{\Phi^{+}} = \frac{1}{\sqrt{2}} (\ket {\uparrow\uparrow} +
\ket{\downarrow\downarrow}$). Since $\sigma$ is an attempted Bell
state preparation, $O_1$ measures the fidelity. The Bell state can be
written in terms of the measurement operators,
\begin{align}
  \ket{\Phi^+ } \bra{\Phi^+} &= \Pi_0 + \Pi_2 \nonumber \\
  &+ \left[U(\tfrac{\pi}{2},0)^{\otimes 2}\right]^{\dagger} \Pi_1 U(\tfrac{\pi}{2},0)^{\otimes 2} \nonumber \\
  &- \left[U(\tfrac{\pi}{2},\tfrac{\pi}{2})^{\otimes
      2}\right]^{\dagger} \Pi_1
  U(\tfrac{\pi}{2},\tfrac{\pi}{2})^{\otimes 2}.
\end{align}
In principle, one could calculate the fidelity by finding the
probability of each measurement outcome with an operator in the
expansion above. However, the SDP algorithm allows for exact
calculation without knowing the expansion.

An example of a set-identifiable observable about which we can gain
partial information is the probability that the second ion is in the
bright state, $O_2 = \ket{\downarrow \uparrow} \bra{\downarrow
  \uparrow}$.  This operator does not have an expansion in terms of
the engineered measurements, unlike the Bell state. However, we can
still gain partial information about the second ion bright via the SDP.
Upper and lower bounds of both example observables are reported in
Table~\ref{tbl:examples} along with the true value of the expectation
value and the value with respect to the point estimate.

\begin{table}
  \centering
  \begin{tabular}{|l|r|r|}
    \hline
    Observable     & $O_1=\ket{\Phi^+} \bra{\Phi^+} $& $O_2=\ket{\downarrow \uparrow} \bra{ \downarrow \uparrow}$ \\ \hline
    True $\langle O \rangle$ & 0.9925 & $2.5\times10^{-3}$   \\ \hline
    M.L. $\langle O \rangle$ & 0.9902 & $2.452\times10^{-3}$   \\ \hline
    SDPs            & (0.9902, 0.9902) & $(2.361,  2.543)\times10^{-3}$   \\ \hline
    Basic C.I.         & (0.9849, 0.9943)  &  $(1.296,  3.451)\times10^{-3}$  \\ \hline
    Bias corrected C.I.           & (0.9841, 0.9944) & $(2.522, 4.357)\times10^{-3}$    \\ \hline
  \end{tabular}
  \caption{ \label{tbl:examples} Numerical results from the two-ion
    example. The procedure is applied with measurement operators and
    high-fidelity unitary processes described in Sec. \ref{sec:alg}
    and unknown state
    $\sigma = 0.99 \ket{\Phi^+ }\bra{ \Phi^+} + \tfrac{0.01}{4}
    \mathds{1}$. ``True $\langle O \rangle$'' is the true expectation
    value for the prepared state, ``M.L. $\langle O \rangle$'' is the
    estimated expectation value with $\hat{\sigma}$, ``SDPs'' are the
    (lower, upper) bounds returned from the semidefinite programs,
    ``Basic C.I.'' is the 95~\% confidence interval found with the
    basic method and ``Bias corrected C.I.'' is the 95~\% confidence
    interval found with the bias corrected method.}
\end{table}

\section{Uncertainties\label{sec:uncertainties}}
To quantify the uncertainty in our procedure we use parametric
bootstrap resampling. For this method, one generates samples from the
estimated probability distribution, which simulates new repetitions of
the experiments, and uses the results to estimate the distribution of
various parameters~\cite{Boos03}. For our case, the estimated
probability distribution is defined by Eq.~\eqref{total_prob} with
$\boldsymbol{\tau} = ( \rho_0, \hat{\sigma})$ and $Q = \hat{Q}$, the
estimates produced by the MLE algorithm. We numerically sample this
distribution $n$ times for each combination of state (index $j$) and
engineered POVM (index $i$) to create a simulated histogram for each
of the experiments described in Sec.~\ref{sec:General_model}. The
bootstrap resampling is performed with coarse-grained measurement
outcomes according to the same binning rule so that the bootstrap
histogram matrices have the same structure as the original binned
experimental histogram matrices.  We repeat the numerical technique
from Sec.~\ref{sec:alg} and expectation value bounding SDPs from
Sec.~\ref{sec:expectation_values} on the numerically generated data to
produce new estimates for the measurements, states, and upper and
lower bounds for $\langle O \rangle$. We repeat this entire bootstrap
procedure $t$ times, producing a distribution of estimated measurement
operators, states, and bounds on observables.

We use the bootstrap distribution to estimate the uncertainty in our
method by calculating approximate $95~\%$ bootstrap confidence
intervals of the bounds on observables. For identifiable observables
(such as the Bell-state fidelity) the confidence interval can be
calculated with standard techniques (see below). However, when
reporting uncertainty for expectation values of set-identifiable
observables a complication arises. In these cases, the expectation
values are lower and upper bounded by the SDPs, but in each bootstrap
resample different lower and upper bounds will be
estimated. Therefore, there exists uncertainty from the observable
being set-identifiable as well as from the bootstrap distribution. To
incorporate both uncertainties, we calculate 97.5~\% one-sided
confidence intervals for the lower and upper bounds separately and
report the overlapping range, which is a 95~\% confidence interval.

There are many methods to determine the bootstrap confidence
intervals, but these methods make assumptions that are often not
satisfied in our situation. In simulations, we see that the bootstrap
distributions for the upper and lower bounds of set-identifiable
observables are commonly biased or asymmetric. This is detected by
observing a difference between the median of the bootstrap resamples
and the estimate from the original data. Bias in maximum likelihood
quantum state tomography has also been reported in
Refs.~\cite{Sugiyama12, Schwemmer15, Silva17}. These effects are caused
by the nonlinearity of the estimator and the complicated structure of
the boundary of quantum state space, which necessarily affects our
lower and upper bounds on expectation values due to the nature of the
SDPs.  Certain techniques to construct confidence intervals, such as
the percentile method~\cite{Efron93}, are sensitive to bias and
asymmetry, and are therefore contraindicated.  Further, the bias and
boundary issues imply that the theory underlying other bootstrap
confidence interval methods is not applicable. As a result, we expect
systematic coverage probability errors that cannot be removed by
increasing the number of bootstrap samples.
  
Nevertheless, for moderate confidence levels between $60~\%$ and
$95~\%$, intervals obtained can still be useful for descriptive
purposes.  To this end, we report two methods, the
basic~\cite{Davison97} and bias corrected~\cite{Efron93} bootstraps,
which both have some robustness to bias and asymmetry, though neither
is designed for high dimensional estimates with boundary
constraints. These methods return sometimes very different confidence
intervals (cf. in Table~\ref{tbl:examples} for $O_2$). Therefore, we
stress that both should be taken only as qualitative descriptions of
the uncertainty and not used for further inference.

More sophisticated methods to deal with bias and asymmetry exist
(for example, the bias corrected and accelerated method in
Ref.~\cite{Efron93}).  However, such methods are not designed to
address the underlying problems encountered. Another possibility may
be to use methods that involve a double bootstrap, but they are
impractical for our procedure.  Further research is required to obtain
high-quality bootstrap confidence intervals for quantum tomography.
Alternatively, one might adapt the confidence regions described in
Refs. \cite{Christandl11, Blume-Kohout12}, which are not based on the
bootstrap.

The bootstrap procedure also allows us to perform a likelihood ratio
test to determine how well our model fits the observed data. The
``likelihood ratio'' is the ratio of the likelihood of our null model,
the estimates of $\hat{\sigma}$ and $\hat{\boldsymbol{F}}$ from the
experimental data (with log-likelihood given by the iterative program
discussed in Sec.~\ref{sec:alg}), to the likelihood of an alternative
model, the experimental frequency histogram matrix with log-likelihood
function,
\begin{equation}
  \lik_{\textrm{frq}}(\boldsymbol{H}) = \sum_{i,j,c} H_{i,c}^{(j)} \ln H_{i,c}^{(j)}.
\end{equation}
To perform the likelihood ratio test we compare the likelihood ratio
of the original estimates to the distribution of likelihood ratios
created by the bootstrap procedure.  The likelihood ratio test
statistic is
\begin{equation}
  \Lambda_0= \lik(\hat{\sigma}, \hat{\boldsymbol{F}} |\boldsymbol{H}, \boldsymbol{\mathcal{U}},\rho_0) - \lik_{\textrm{frq}}
  (\boldsymbol{H}).
\end{equation}
We also compute an analogous statistic for each of the $t$
bootstrapped data sets (each computed using its own simulated
histogram matrices and respective state and measurement estimate) to
generate a distribution of likelihood test statistics, $\{
\Lambda_i|i=1,\dots,t \}$.  Because the bootstrap data sets are
certainly well described by the null model, their likelihood ratios
should typically be comparable to the likelihood ratio of the
experimental data set, which may or may not obey the model. A
statistically significant difference between the experimental data
set's likelihood ratio and the bootstrap data sets' is evidence that
the null model does not match the experiment. To quantify evidence
against the null model, we compute an empirical p-value by determining
the percentile at which $\Lambda_0$ falls in the distribution from the
bootstrap data sets~\cite{Boos03}.

\section{Conclusion} \label{sec:conclusions} We have developed a
procedure to simultaneously characterize unknown quantum measurements
and states by using a limited set of high-fidelity quantum
operations. The protocol requires two types of experiments, which we
designated as ``reference'' and ``probing.'' The reference experiments
use the high-fidelity processes to estimate the unknown measurement
operators, similar to QDT. The probing experiments use the
high-fidelity processes to probe unknown state preparations, similar
to QST. In our procedure the estimation of the measurement operators
and density matrices is achieved simultaneously via alternating
MLE. This means the estimates produced are consistent with both
reference and probe experiments. We also introduced a method for
estimating expectation values of the unknown states by two SDPs when
the high-fidelity processes do not produce IC measurements.

Our procedure applies to systems where we can apply certain operations
with high fidelity.  These need not be universal; a small set of
single-qubit unitaries suffices. A sufficient set of such operations
is available in many state-of-the-art quantum information processors,
for example trapped ions. Moreover, our protocol has the advantage
that it is efficient to implement relative to previous proposals. This
is because we make use of prior information about the measurement
operators and do not seek to diagnose all parts of the quantum
system. However, since our protocol is dependent on prior information
about the system, it is not ideal for experiments that have not
previously been diagnosed.  In the absence of a well defined initial
state and sufficiently many high-fidelity operations, methods such as
GST may be better suited.

Our implementation and testing of this estimation procedure has
focused on a few trapped ions. In this system, measurements can be
modeled as classical noise following projection onto a small number of
orthogonal subspaces and measurement outcomes that can be partitioned
into a small number of bins with little information loss. In
principle, the procedure can handle higher-dimensional systems and
measurements that consist of multi-dimensional histograms but further
optimization is required to run these cases efficiently. There are
also opportunities for expanding our procedure by applying it to
systems with measurements that cannot be modeled as orthogonal
projection followed by classical noise.

\appendix
\section{Software implementation\label{app:software}} 
We implemented our procedure as a Python package, available at
Ref.~\cite{state_meas_tomo}. Instructions for installing and running
the package are given in the accompanying documentation.  Our package
contains a version of $R \rho R$ for the first optimization
subproblem, and uses \texttt{scipy.optimize} for the second
optimization subproblem. The SDPs that estimate the expectation values
described in Sec.~\ref{sec:expectation_values} require the MATLAB API
engine. The SDPs are solved with YALMIP~\cite{Lofberg2004}, which is a
package for MATLAB.

\section{Coarse-graining measurement outcomes} \label{app:binning}
The measurement device may have a very large number of possible
outcomes $b = 1, \dots, M$ where $M \gg 1$.  This is a main
contributor to the computational complexity of the log-likelihood
maximization, because each outcome adds a term to the log-likelihood
function, which requires more computation for finding $R$ in the first
subproblem and increases the optimization space for the second
subproblem.  To reduce the complexity, we coarse grain the outcome
space by collecting the original outcomes $b$ into $G < M$ bins. (The
original outcomes can also be thought of as binned, so this procedure
is technically a re-binning.) We assume that the ordering of the
original outcomes $1,\ldots,M$ is meaningful so that it makes sense to
bin consecutive outcomes for minimum information loss. The bins are
then identified by the bin edges $B_c$ in the outcome space
$\{B_c |c = 0,\dots, G \}$, where $B_0=0$, and $B_{G} = M$, so an
outcome $b \in (B_{c-1},B_{c}]$ is mapped to the coarse-grained
outcome $c$.  Coarse graining reduces the information about the
unknown states and measurements that was captured by our
experiments. For an extreme example choose $G = 2$ with $B_1 = 0$;
then every outcome $b$ is mapped to the coarse-grained outcome $c=1$,
which provides no information. To combat this problem, we construct a
heuristic algorithm to choose the bin edges. We apply this algorithm to
``training data,'' which is composed of 10~\% of the trials from the
reference experiments randomly sampled without replacement and set
aside from all further analysis. Though we describe the algorithm in
terms of one-dimensional histograms here, our software also supports
multidimensional histograms.

For a given number $G$ of coarse-grained bins, our algorithm maximizes
the amount of information retained in the coarse-grained outcomes $c$
by a heuristic based on mutual information. First, we need to identify
the information about the unknown states and measurement. In our
procedure, this information is contained in the histograms collected
from the experiments. These histogram matrices $H^{(j)}_{i,b}$ are
rectangular with vertical dimension $M$, the number of possible
outcomes. Coarse graining the outcomes then shrinks the vertical
dimension to $G < M$. For bin edges $\{ B_c \}_c$ the coarse grained
histogram matrix is defined as
\begin{equation} \label{H_bin}
  H'^{(j)}_{i,c}=\sum_{b=B_{c-1}+1}^{B_{c}}H^{(j)}_{i,b}.
\end{equation}
Note that every histogram (that is, every row of the histogram matrix) is
treated identically. The coarse-graining is also applied uniformly to
the transition matrix
\begin{equation} \label{Q_bin}
  Q'_{k,c}=\sum_{b=B_{c-1}+1}^{B_{c}}Q_{k,b}.
\end{equation}
Therefore, both $H'^{(j)}$ and $Q'$ have smaller vertical dimension
due to the coarse graining. In the following, we study the
coarse-grained histogram matrices and transfer matrix estimated from
the training data $H^{\textrm{train}} = H'^{(0)}$ and
$\hat{Q}^{\textrm{train}} = Q'$ from Eq.~\eqref{q_est}.

We quantify the amount of information retained in coarse-grained
histograms with the mutual information. Consider a joint probability
distribution $P(k,c)$, where $c$ are the bin indices and $k$ are
underlying outcomes. Let $C$ and $K$ denote the random variables with
values $c$ and $k$, respectively.  The mutual information $I_{P}(K;C)$
between $K$ and $C$ quantifies the amount of information $C$ has about
$K$ (or $K$ about $C$). It is given by
\begin{equation}
  \label{mutual_info}
  I(K; C) = \sum_{k,c} P(c)P(k|c)
  \log_2\left( \frac{P(k |c)}{P(k)} \right),
\end{equation}
where $P(k |c)=P(k,c)/P(k)$ is the conditional probability
distribution.  Our goal for coarse graining is to maximize the mutual
information, so as to approach the full information in $K$, achieved
when $C=K$. We could calculate the mutual information between the
original and the coarse-grained histograms, but what we really care
about is how well the coarse-graining retains information about the
underlying outcome distribution $p_k = \Tr{(\Pi_k \rho)}$ discussed in
Sec.~\ref{sec:General_model}. This distribution is dependent on the
state, so we choose a representative state $\rho= \rho^{\textrm{eq}}$
with the property that underlying outcomes have equal probability,
namely $P(k) = \Tr{(\Pi_k \rho^{\textrm{eq}})} = 1/N$ for all $k$. We
could estimate the coarse-grained histogram matrix $H^{\textrm{eq}}$
for $\rho^{\textrm{eq}}$ based on the training data's histogram, but
this relation is dependent on the family of known input states
$\boldsymbol{\rho}$. Instead we find it more convenient to determine
$H^{\textrm{eq}}$ based on the estimated transfer matrix
$\hat{Q}^{\textrm{train}}$, which already contains information about the
histogram and the known states, such that $H^{\textrm{eq}}_c = \sum_k
\hat{Q}^{\textrm{train}}_{k,c} \tfrac{1}{N} = P(c)$.  The needed joint
probability distribution is given by $P(k,c)= P(c|k)P(k)=P(c|k)/N$
with $P(c|k)$ given by $\hat{Q}^{\text{train}}_{k,c}$.

\begin{figure}
  \includegraphics[width=3.4in]{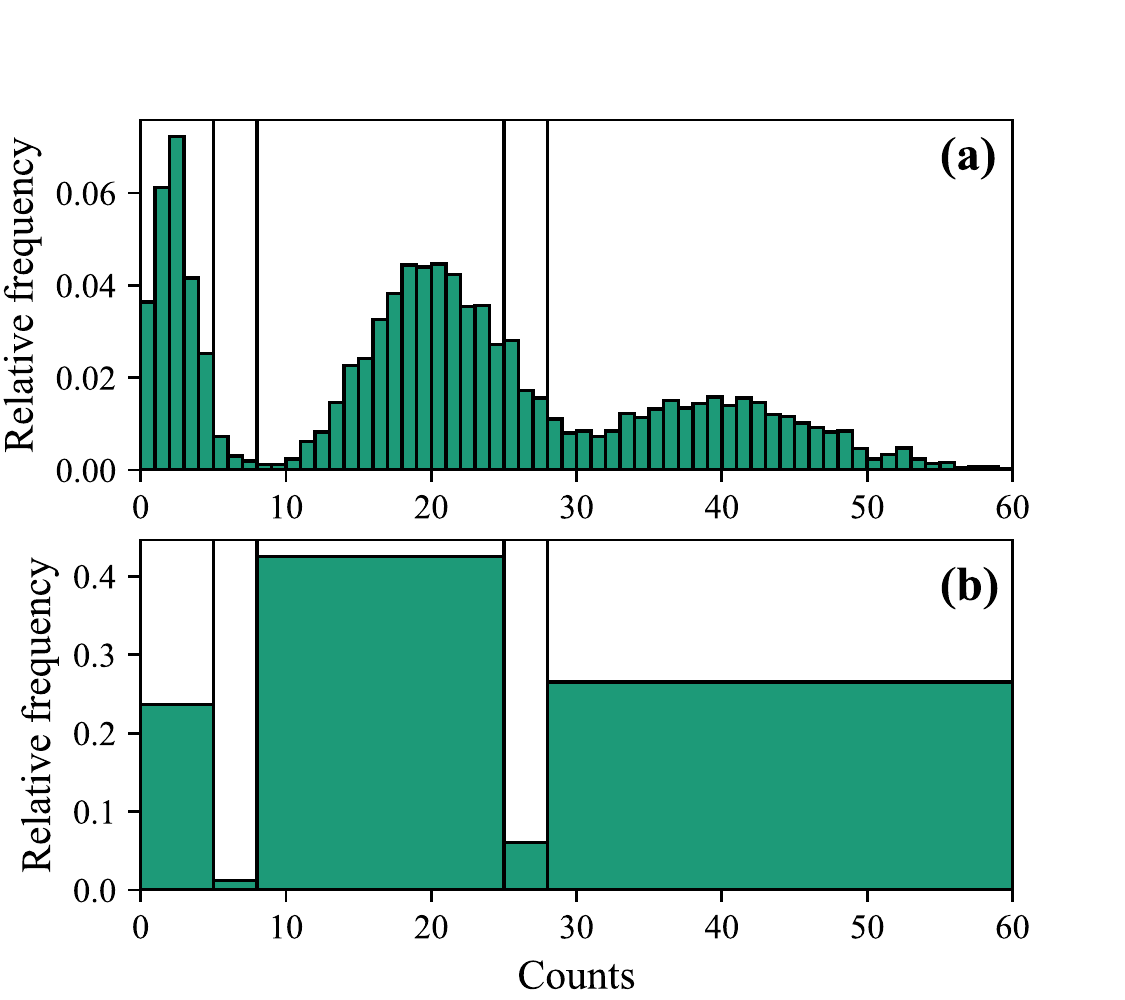}
  \caption{\label{binning}Illustration of the binning procedure. (a)
    Relative frequency histogram for the second reference experiment,
    $U_1 = U(\tfrac{\pi}{2},0)^{\otimes 2}$. (b) Same histogram data
    but after binning. The black lines show the bin boundaries found
    with the heuristic algorithm.}
\end{figure}

We determine a good coarse graining by finding bin edges $\{ B_c \}_c$
with high mutual information for $P(k,c)$ as defined in the previous
paragraph. Searching over all possible bin edges to maximize $I(K;C)$
is impractical, so instead we use a heuristic algorithm that
iteratively adds bin edges. We pre-specify the target number of bins
$G$. Then, starting with two bins, we compute the mutual information
for each possible location for the bin edge between $0$ and $M$.  The
boundary location that gives the largest mutual information is then
fixed as an element of what will be our final list of bin edges. To
add a third bin, leaving the existing bin edge fixed, we compute the
mutual information for all possible locations for the new edge. The
new edge that gives the largest mutual information is added to
$\{ B_c \}_c$. We continue this procedure until the target number of
$G$ bins is reached. We apply the resulting bin edge rule to the
remaining reference histograms not used in the training data, all
probing histograms, and $\hat{Q}^{\text{ini}}$ by Eqs.~\eqref{H_bin}
and \eqref{Q_bin}. An example of binning ion fluorescence data is in
Fig.~\ref{binning}.

We have observed that binning consecutive elements in the rows of
$H^{(j)}$ has been effective when systems occupying a single subspace
produce unimodal distributions, such as those produced by the ion
measurements. Analysis of multimodal distributions may benefit from
more complicated binning strategies.

\section{Stopping criteria \label{app:stopping}}
After each optimization subproblem is run in an iteration, we bound
the difference between the current log-likelihoods $\lik_i$ ($i=1$ and
$i=2$ for the first and second subproblems) and their respective
maximum possible values. The bounds tell us how much the
log-likelihoods could increase with further iterations. We stop the
algorithm when both differences are individually below pre-specified
thresholds $T_{\sigma}$ and $T_{Q}$.

To find the difference bound for the first subproblem, we follow the
method introduced in Ref.~\cite{Glancy12}.  The bound
$\hat{S}_{\sigma}$ is given by
\begin{equation}
  \lik_1(\sigma_{\textrm{ML}}) - \lik_1(\hat{\sigma}^{\text{cur}}) \leq \max \{ \eig [ R(\hat{\sigma}^{\text{cur}}) ]\} - n = \hat{S}
  _{\sigma},
\end{equation}
where $\sigma_{\textrm{ML}}$ is the maximum likelihood state and
$\max \{ \eig [ R(\hat{\sigma}^{\text{cur}}) ] \}$ is the maximum
eigenvalue of $R(\hat{\sigma}^{\text{cur}})$, which was computed by
the $R \rho R$ algorithm. We stop iterations of $R \rho R$ when
$\hat{S}_{\sigma} \leq T_{\sigma}$.

The difference bound for the second subproblem is calculated in a
similar way. Since the log-likelihood function of this subproblem is
also concave, the difference between the maximum log-likelihood and the
log-likelihood of $\hat{Q}^{\text{cur}}$ is upper bounded by
\begin{equation} \label{Q_bound} \lik_{2}(Q_{\textrm{ML}}) -
  \lik_{2}(\hat{Q}^{\text{cur}}) \leq \sum_{k,c}
  (\hat{Q}^{\text{cur}}_{k,c} - Q^{\textrm{ML}}_{k,c}) \frac{\partial
    \lik_{2}(\hat{Q}^{\text{cur}})}{\partial Q_{k,c}},
\end{equation}
where $\frac{\partial \lik_{2}(\hat{Q}^{\text{cur}})}{\partial
  Q_{k,c}}$ are the elements of the gradient of $\lik_{2}$. Since we
do not know $Q_{\textrm{ML}}$, we bound the right side of
Eq.~\eqref{Q_bound} by finding the maximum value over all
distributions by solving the optimization problem
\begin{equation}
  \begin{aligned}
    & \underset{X}{\text{maximize:}} & & \sum_{k,c}
    (\hat{Q}^{\text{cur}}_{k,c} - X_{k,c}) \frac{\partial
      \lik_{2}(\hat{Q}^{\text{cur}})}{\partial \hat{Q}^{\text{cur}}
      _{k,c}}, \\
    & \text{subject to:}
    & & \sum_c X_{k,c} = 1 , \; \forall k, \\
    &&& 0 \leq X_{k,c} \leq 1, \; \forall \; k,c.
  \end{aligned}
\end{equation}
The value $\hat{S}_Q$ returned is an upper bound on the difference
between the maximum log-likelihood and the log-likelihood at the current
iteration.

We compare both bounds to the pre-specified thresholds $T_{\sigma}$
and $T_{Q}$. Default values for these thresholds in the code are
currently $T_{\sigma}=0.3$ and $T_{Q}=0.25$, which were chosen
empirically by testing on simulated data representative of our
ion-trap applications. When $\hat{S}_{\sigma} \leq T_{\sigma}$ and
$\hat{S}_{Q} \leq T_{Q}$ the procedure is stopped, and the final
estimates are returned.

\begin{acknowledgments}
  This work includes contributions of the National Institute of
  Standards and Technology, which are not subject to
  U.S. copyright. The use of trade names allows the calculations to be
  appropriately interpreted and does not imply endorsement by the US
  government, nor does it imply these are necessarily the best
  available for the purpose used here.
\end{acknowledgments}

\bibliography{bib} 


\end{document}